\documentclass[jcp,twocolumn,english,superscriptaddress]{revtex4-1}

\usepackage[T1]{fontenc}
\usepackage[latin9]{inputenc}
\usepackage{color}
\usepackage{babel}
\usepackage{latexsym}
\usepackage{float}
\usepackage{amsmath}
\usepackage{amsfonts}
\usepackage{graphicx}
\usepackage{times}   %% Times Roman font
\usepackage{esint}
\usepackage{subfigure}
\usepackage{verbatim}
\usepackage[unicode=true,pdfusetitle,
 bookmarks=false,colorlinks=true,citecolor=blue,urlcolor=blue,linkcolor=red]{hyperref}

\makeatletter

\@ifundefined{textcolor}{}
{%
 \definecolor{BLACK}{gray}{0}
 \definecolor{WHITE}{gray}{1}
 \definecolor{RED}{rgb}{1,0,0}
 \definecolor{GREEN}{rgb}{0,1,0}
 \definecolor{BLUE}{rgb}{0,0,1}
 \definecolor{CYAN}{cmyk}{1,0,0,0}
 \definecolor{MAGENTA}{cmyk}{0,1,0,0}
 \definecolor{YELLOW}{cmyk}{0,0,1,0}
}

\makeatother

\newcommand{\SAVE}[1]{}

\newcommand{\prlsec}[1]{\emph{#1---}}

\begin{document}
\renewcommand\abstractname{}

\title{Measuring quantum entanglement, machine learning and wave function tomography: Bridging theory and experiment with the quantum gas microscope }
\author{Norm M. Tubman}
\affiliation{Department of Chemistry, University of California, Berkeley, Berkeley, California 94720, USA}
\date{\today}
\begin{abstract} 
There is an enormous amount of information that can be extracted from the data of a quantum gas microscope that has yet to be fully explored.   The quantum gas microscope has been used to directly measure magnetic order, dynamic correlations, Pauli blocking, and many other physical phenomena in several recent groundbreaking experiments.   %However, the full potential of a quantum gas microscope has not yet been fully realized.  
However, the analysis of  the data from a quantum gas microscope can be pushed much further, and when used in conjunction with theoretical constructs it is possible to measure virtually any observable of interest in a wide range of systems.  We focus on how to measure quantum entanglement in large interacting quantum systems.  In particular, we show that quantum gas microscopes can be used to measure the entanglement of interacting boson systems exactly, where previously it had been thought this was only possible for non-interacting systems. We consider algorithms that can work for large experimental data sets which are similar to theoretical variational Monte Carlo techniques, and more data limited sets using properties of correlation functions.

\end{abstract}

\maketitle

\newpage

\prlsec{Introduction} 
 Cold atom experiments provide a platform in which the many body physics of explicitly set Hamiltonians can be measured directly.  One experimental device used in cold atom experiments is the quantum gas microscope.  Direct measurement of magnetic order and dynamic correlations is feasible with such experimental systems~\cite{Bakr2009,Sherson2010,gasexp-1,gasexp-2,gasexp-3,Omran2015,Preiss2015,Trotzky2010}.  However, the full information provided from a quantum gas microscope is not limited to simple real space observables.   A reconstruction of a bosonic quantum wave function is not only possible, but  tractable.  A similar reconstruction is feasible for a systems of fermions, but with a few limitations (as will be discusssed).  For a low temperature gapped system (and all zero temperature systems) the distribution of configurations generated from a quantum gas microscope is given by $|\langle \psi_{gs} | \mathbf{R} \rangle| ^{2}$, and for finite temperature system the distribution is given by the diagonal elements of the finite temperature density matrix~\cite{Ceperley1995}. %$\langle \mathbf{R}| \rho | \mathbf{R} \rangle $~\cite{Ceperley1995}.  
 %How to use this information to generate correlation functions and order parameters that are of general interest is the primary focus of this work. 
   Other types of wave function tomography have also been considered previously~\cite{Lundeen2011,Swingle2014,Baumgratz2013,Cramer2010}, and the techniques considered here for the quantum gas microscope should be considered an alternative, especially for cases in which one is interested in many-body systems. 
The ideas presented here are adapted from the long standing and widely used literature of quantum Monte Carlo research as well as new ideas from the machine learning community~\cite{Tubman2014,Tubman2012,McMinis2013,Toulouse2007,Melko2016-1,Melko2016-2,Melko2016-3,Bajdich2010}.  %The information typically used by quantum Monte Carlo practitioners is the same information that is generated by a quantum gas microscope. 
 We are able to describe in this work how measurement of difficult quantities, such as real space quantum entanglement, can be made.  This process blends theory and experiment providing new insights in both directions.

\prlsec{From Experiment to Theory}
In this section we consider how to expand the current analysis of quantum gas microscope data and develop theoretical constructs for which we can develop detailed information about quantum systems.    We consider two types of wave function tomography, which is also known in the literature as quantum state tomography.  % that can be applied.  
The methods we present include a general approach and an approach for situations in which the amount of data that can be generated is limited. 
%and also cases of in which one wants to study homogenous or inhomogenous phases.

\begin{figure}[htpb]
\centering
\includegraphics[width=0.49\linewidth]{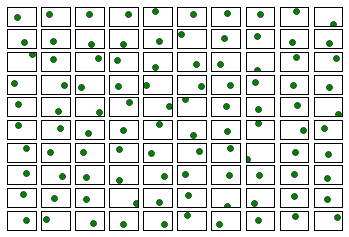}
\includegraphics[width=0.49\linewidth]{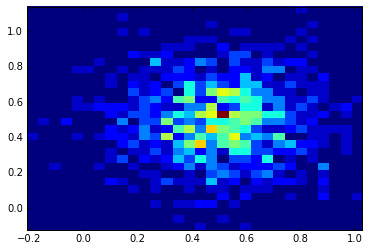}
\includegraphics[width=0.49\linewidth]{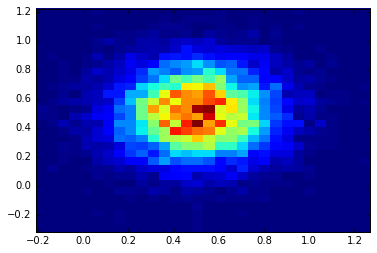}
\includegraphics[width=0.49\linewidth]{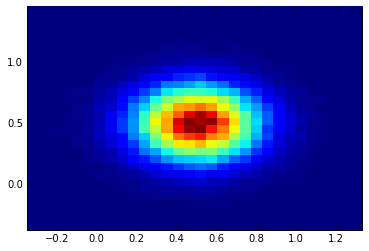}
\includegraphics[width=0.49\linewidth]{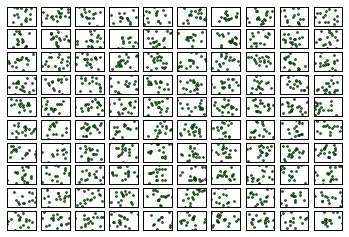}
\caption{Top and middle rows:  An example of a single particle reconstruction from snapshots of a quantum particle in a 2D harmonic trap.  The top left figure is an example of data that might be generated from 100 snapshots, and the other three figures show what an example reconstruction, based on a 2D grid, would be with 1,000, 10,000, and 100,000 snapshots.  The units of distance are arbitrary.  Bottom row: A set of data that would could be generated when multiple particles are imaged.  When proper techniques are used the efficiency of determining a quantum state is enhanced such that even the high-dimensional multi-particle state can be reconstructed.  
}
\label{fig:recon} 
\end{figure}

\prlsec{Wave function tomography}
In the limit of infinite samples from a quantum gas microscope, it is possible to recreate the full bosonic  or Fermionic wave functions without approximation under very general conditions.  In Fig.~\ref{fig:recon} we have illustrated how a single particle wave function might be reconstructed from varying number of snapshots, and what this implies for many particle systems.   This process is limited by the number of samples that can be generated experimentally and the number of variational parameters that we can reasonably optimize; This is directly analogous to optimizing wave function with variational Monte Carlo~\cite{Foulkes2001}.  Generally quite complicated wave function forms can be optimized~\cite{Clark2011,Yubo2015,Mcmillan1964,Bajdich2008,Neusman2013,Drummond2006,Mitroy2013,dmrg,Ortiz1993,Jones1997}.

  %Although a quantum gas microscope does not generate phase information for quantum configurations, it is possible to use the same techniques described in this section with Fermions.  However, because the phase information is not present, some of the extended techniques that work for data limited experiments, described below, will not work for Fermions without approximation.  

To develop practical tools we consider the variational Monte Carlo framework which can be applied to \textit{ab initio} systems and lattice simulations~\cite{Foulkes2001,Shap2016,Clark2011-1,Tubman2012}.  Real space variational Monte Carlo is based on generating configurations in real space from a trial wave function.  These configurations are used to evaluate a cost function, that often involves the system Hamiltonian, from which a set of variational parameters are optimized.  There are several widely used optimization techniques including variance and energy optimizations~\cite{Toulouse2007}, as well as wave function overlap optimization~\cite{Bajdich2010}.  Since initial configurations are generated from an approximate guess wave function,  variational Monte Carlo proceeds in iterations, during which new configurations are generated from the current best wave function.  % in which new configurations are generated from the current best wave function, after which a new optimization cycle begins.    

The main idea in this work is that the steps to generate new configurations in variational Monte Carlo can be replaced by the data generated from a quantum gas microscope.
  The approach is different from standard variational Monte Carlo as the generated snapshots from the quantum gas microscope are directly sampled from the wave function of interest and therefore there is no need to run several iterations.      
The question then becomes what type of variational parameters can be optimized and how many snapshots are needed.   We discuss below the large data limit and the small data limit.  For the large data limit, there is no difference applying the technique to bosons or Fermions,  except that an antisymmetric variational wave functions must be used for systems of fermions.   We limit the discussion mainly to zero temperature/low temperature experiments except for some comments at the end about dealing with density matrices instead of wave functions.  

\prlsec{Large data limit}
In the large data limit, one can proceed to run wave function optimization routines, as discussed above,  on a set of snapshots generated by a quantum gas microscope.  There are many available libraries and codes in which these optimizations can be done, however some expertise is often helpful.
There are several strategies that that one can use for selecting variational parameters for continuum and lattice Hamiltonians.  In our previous quantum Monte Carlo work we directly make use of wave function optimization~\cite{Tubman2011,Tubman2014-1,Tubman2012,Tubman2012-1,Tubman2014,Tubman2015,Tubman2015-1,Yubo2015,Tubman2016}. With our variational Monte Carlo tools we allow for the optimization thousands of variational parameters~\cite{McMinis2013,Swingle2013,Tubman2014,Yubo2015}.  For bosons we generally only optimize two-body and three-body Jastrow terms, which are suitable for a wide range of systems, as will be discussed in relation to the no-nodes theorem~\cite{Ceperley1995}.  
The form of a Jastrow wave function can be written as $\Psi(\mathbf{X}) = e^{J(\mathbf{X})}$ for a system of bosons and $\Psi(\mathbf{X}) = e^{J(\mathbf{X})}D(\mathbf{X})$ for a system of fermions. For these wave function,  $J$ is the Jastrow and $D$ is an antisymmetric function of the particle coordinates. The coordinates $\mathbf{X} = (\mathbf{x_{1}},\mathbf{x_{2}},...,\mathbf{x_{n}})$, and $\mathbf{x_{i}} = (\mathbf{r_{i}},\sigma_{i})$ is the space and spin coordinates of a single particle.
Often one and two-body Jastrow terms are used which is written as,
\begin{equation}
J(\mathbf{X}) = \sum_{i}^{N}\xi(\mathbf{x_{i}}) - \frac{1}{2}\sum_{i=1}^{N}\sum_{j=1,j\neq i}^{N}u(\mathbf{x_{i},x_{j}})
\end{equation}
The one-body Jastrow, $\xi$, is generally a set of functions centered around fixed nuclei or lattice sites.  A two-body Jastrow, $u$, consists of pairwise terms between all quantum particles~\cite{Foulkes2001,Mcmillan1964}.  These functions depend only on the distance between quantum particles, and in our variational Monte Carlo studies use on the order of 10-20 variational parameters that are the knots of a spline function in real space.  Three-body Jastrow terms can be much more complicated and depend on distances and angles between all triplets of quantum particles.  However these can be fully optimized as well.

For Fermions we generally optimize more complicated forms, that involve backflow and multi-determinants~\cite{Clark2011,Yubo2015,Drummond2006}.   In systems that are homogenous for which the Jastrows factors can be replicated for all lattice sites or atoms, there might be only a few dozen parameters to optimize. Generally only a few hundred configurations are needed in such simulations (or less depending on the situation)  to optimize the variational parameters.   For Fermions we generally try to include many more variational parameters through multi-determinant expansions.  When optimizing a thousand or more variational parameters we generally try to include 10,000-100,000 snapshots for the optimizations.  The need for many configurations comes not only from the the number of variational parameters we are trying to optimize, but also that the determinant parameters in a determinant expansion become very small, and require many samples in order to resolve their values above the inherent noise.  

As mentioned previously, other types of variational parameters can be included in an optimization.  In principle even a tensor network~\cite{dmrg,dmrg_review,ITensor,Stoudenmire_White_2D_DMRG,dmrg_white} would be an alternative form that can be optimized, although it would require expanding on the techniques discussed here.    We also mention here, in connection with the machine learning techniques in later sections, that it is theoretical possible to optimize variational parameters of a neural network to learn a quantum wave function, both with variational Monte Carlo and a quantum gas microscope.  Advances with symmetry functions have made it feasible to learn very high dimensional potential energy landscapes~\cite{nong2011,nong2016}, and we suggest here that it is possible to use the same symmetry functions in order to have a neural network learn a wave function of bosons.  While in principle it is possible to learn Fermionic wave functions with neural networks,  enforcing the antisymmetry requires some additional developments.
 
\prlsec{Small data limit for Bosons}
Alternatives exist for data limited experiments beyond the full wave function tomography discussed in the previous section.  %Such approaches can also benefit from working on highly homogenous systems and systems with many particles.
 The full power of a quantum gas microscope on a system of bosons can be best understood from the Feynman "no node" theorem.  This is a statement about how a boson ground state quantum wave function in coordinate representation is positive definite~\cite{Feynman1972,Wu2009}.  Feynman was able to show that this holds under a broad set of conditions which includes that no external rotations be applied.  Because of time reversal symmetry, the wave function can be chosen to be real, and the "no node" theorem can be shown to follow from a procedure that always lowers the energy of a system by removing nodes.  Virtually all boson ground states that do not involve applied magnetic fields can realize a wave function with no nodes which includes many exotic states.    
However, only for bosons does the wave function amplitudes and wave function probability distribution have identical information.  

The probability distribution of bosons in coordinate space $|\psi_{exp}(\mathbf{x_{1}...x_{N}})|^{2}$ is what a quantum gas microscope would measure in the limit of an infinite number of snapshots of a quantum system.  In other words a quantum microscope can be used to sample a many-body wave function of interest according to its probability distribution  $|\psi_{exp}|^{2}$ after which  $\psi_{exp}$ can be determined by taking a square root of the probability distribution.   This is not generally true in fermionic systems or complex wave functions, where the sign/phase of a wave function can not be extracted from the probability distribution.
  
  \begin{figure}[htpb]
\centering
\includegraphics[width=0.49\linewidth]{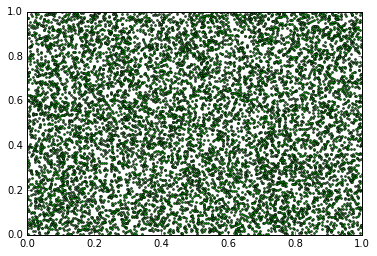}
\includegraphics[width=0.49\linewidth]{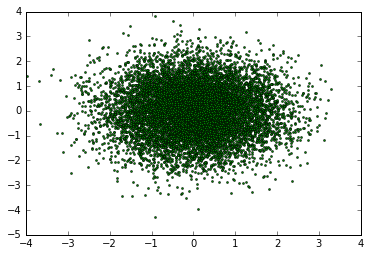}
\includegraphics[width=0.49\linewidth]{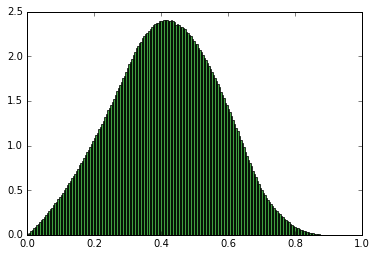}
\includegraphics[width=0.49\linewidth]{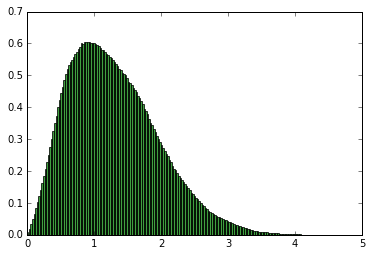}
\caption{Top row:  A single snapshot of 10,000 non-interacting particles (left) uniformly distributed in a 2D box, and (right) in a 2D harmonic trap.  The units of distance is arbitrary.  Middle row:  Normalized  radial correlation function for the snapshots of the systems pictured in the first row.  The radial correlation functions are generated only for pairs in which at least one of the atoms is within 0.3 distance units from the center of these systems.  For these system sizes, already the single particle pair correlation function can be reconstructed with very low noise. 
}
\label{fig:pair} 
\end{figure}
  
In very large systems, as with many cold atom experiments where there can be more than a million quantum particles~\cite{Meldgin2016,Mckay2016,Ray2013},  many snapshots would be needed as more variational degrees of freedom are needed to represent a wave function.   Such complications  are compounded where  there are multiple quantum phases present due to differences between the environment in the middle of a trap versus the edge of the trap~\cite{Ray2013,Mckay2016,Meldgin2016}. 
 Historically there have been different approaches to further study such systems through their correlation functions.     Jastrow factors that involve 1, 2 and 3 body functions are regularly used to either minimize the energy (as discussed in the previous section)  or to match known correlation functions.  Matching correlation functions by optimizing Jastrow variational parameters is one possible approach, but one could also use other techniques such as those related to the hypernetted-chain approximation~\cite{Hansen1976}.   The benefit of correlation matching is that it allows one to focus on the phase of interest with high precision as the low order correlation functions may be determined with only a very few number of snapshots.%The former is what we considered in the previous section, and the later is what we are considering in this section. 
 
\begin{figure}[htpb]
\centering
\includegraphics[width=0.49\linewidth]{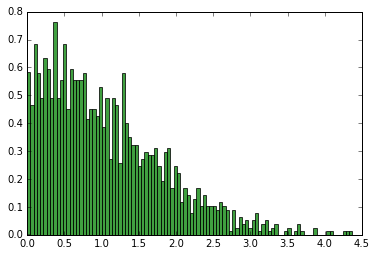}
\includegraphics[width=0.49\linewidth]{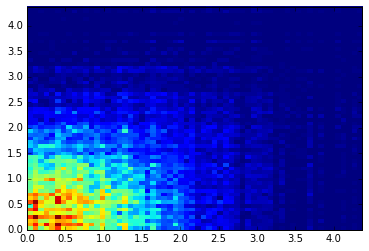}
\includegraphics[width=0.49\linewidth]{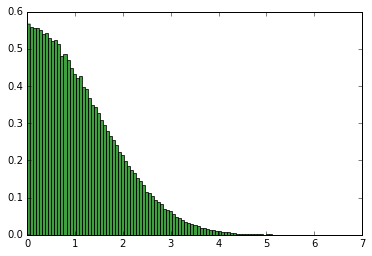}
\includegraphics[width=0.49\linewidth]{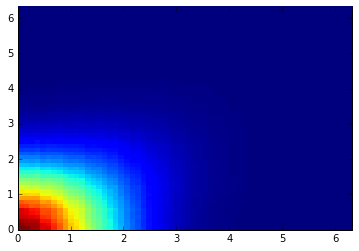}
\caption{Top row:  The two and three particle correlation functions for 60 particles in a non-interacting 1D harmonic trap.  The correlation functions are noisy at this resolution.   More snapshots and larger systems can be used to lower the noise further.
 Bottom row: The two and three particle correlation functions for 600 particles in a non-interacting 1D harmonic trap.  The correlation functions are smooth and accurate over the main regions of interest.   
}
\label{fig:noise} 
\end{figure}

  To understand correlation matching, we start by considering a set of samples from a quantum gas microscope, \{S$_{1}$...S$_{N}$\}.    The density can be estimated from the snapshots and should match the corresponding one body correlation function of the wave function given by,
  \begin{equation}
    P(\mathbf{x_{1}}) = N\int D\mathbf{x_{2}}...\mathbf{x_{N}}\Psi_{exp}(\mathbf{x_{1}}...\mathbf{x_{N}})\Psi_{exp}^{*}\mathbf{(x_{1}}..\mathbf{x_{N}})
   \end{equation}.   
    The two body correlation function can be generated by considering the distance between all pairs of quantum particles over all snapshots, as for example in Fig. \ref{fig:pair} and Fig. \ref{fig:noise}.   The corresponding quantity to be matched by a wave function is given by 
   \begin{equation}
   P(\mathbf{x_{1}},\mathbf{x_{2}}) = \frac{N(N-1)}{2}\int D\mathbf{x_{3}}...\mathbf{x_{N}}\Psi_{exp}(\mathbf{x_{1}}..\mathbf{x_{N}})\Psi_{exp}^{*}(\mathbf{x_{1}}...\mathbf{x_{N}}),
   \end{equation}
    %where the Jastrow factor would be optimized to produce the experimental pair correlation function.
       Similarly, the three body correlation function can be tabulated from the experimental data and a Jastrow factor can be optimized.   The optimization should be done simultaneously for all variational parameters and all correlation functions.   
    % $P(x_{1},x_{2},x_{3}) = \int Dx_{4}...x_{N}\Psi_{exp}(x_{1}..x_{N})\Psi_{exp}^{*}(x_{1}..x_{N})$ matches the experimental three-body correlations.  
For an isotropic system (or an isotropic phase in the middle of a trap), the correlation functions can be reduced in dimensionality. The pair correlation function can be turned into the radial distribution function, $P(r_{1}) =  P(\mathbf{|x_{1}}-\mathbf{x_{2}}|)$.   Excellent approximate wave functions have been generated for various quantum systems, with even just a two body Jastrow term~\cite{Mcmillan1964,Ceperley_Alder}.   

To understand the limit in which the correlation function approach becomes exact and when it should be accurate, we can consider the problem of fitting to a very high dimensional N-body Jastrow term.  If accurate N-body correlation data is measured and tabulated for a system, then  an N-body Jastrow term can be computed. In the case of bosons, the N-body Jastrow term is the exact full wave function of the system, up to statistical errors and limitations in the variational parameters.  This is because a nodeless wave function, as previously mentioned, has the property that the probability distribution can be used to generate the wave function amplitudes by  $\sqrt{|\Psi|^2}$.  Thus we expect the accuracy for our wave function to increase and eventually converge by increasing the order of the correlation functions used.

 For a system with a very large homogenous region, it may be possible to fit several low order correlation functions with just a single snapshot as demonstrated in Fig.~\ref{fig:pair} and Fig.~\ref{fig:noise}.  A typical cold atoms system might have a few hundred thousand bosons %of which if even 10,000 are in a region of interest, 
of which a single snapshot includes a number of pairs proportional to $N^{2}$ and triplets proportional to $N^{3}$. For a system of a few thousand particles, this in some situations will generate enough data such that  low order correlation functions may be tabulated with small error bars with only a few snapshots.  The effectiveness of this approach will depend upon the nature of the correlations and the level of inhomogeneity. % We also note that it is possible to fit inhomogenous systems, with spatial dependent Jastrow factors, something which is done routinely in \textit{ab initio} quantum Monte Carlo.

The above procedure does not work for Fermions, unless an antisymmetric wave function is assumed, and only the Jastrow factor is optimized.  This is a widely used approximation in variational Monte Carlo.  % however it is an approximation when applied to quantum gas microscope data.
  For example a determinant of plane-wave orbitals is often assumed for the antisymmetric part of a wave function when studying interacting Fermi-liquids~\cite{McMinis2013,Swingle2013,Ceperley_Alder,Foulkes2001}.  An exception to this is Fermions in 1D as the nodes of a 1D system are known exactly.  %correlation function matching can be applied to 1D Fermions without any approximations. 
 The nodal surface of a system of N Fermions in one dimension has dimensionality for N-1, which is completely determined by antisymmetry of the wave function.  Only in 1D this true, and it implies for the ground state that the nodal surface is defined when Fermions coincide with each other.  Therefore as long as we use for the antisymmetric part of the wave function a term with the exact nodes, which we can do in 1D, then optimizing a Jastrow factor is as exact as the boson correlation matching.  Approximate antisymmetric wave functions in higher dimensions can be used to generate approximate results, which in many cases may be accurate for a system of interest.%Since there are no known violations of this theorem for systems which observe time reversal symmetry, we can take $\phi$ as equal to an anti-symmetric ground state solution of any non-interacting problem (for example a Hartree-Fock solution, or anti-symmetrized planewaves would work), and optimize the Jastrow in the same way we treated the Boson systems.  Since we know the nodes exactly, and they are correctly produced by $\phi$, the Jastrow term is enough to produce the exact Fermion ground state wave function in the limit that an N-body Jastrow term is used, same as with the boson system.  

\prlsec{Quantum Entanglement}
Using the wave function tomography in the previous section provides experimental access to otherwise difficult to measure quantities, which includes the quantum entanglement of many particle systems.  Real space quantum entanglement has proven to be important in many modern theoretical studies of both high energy and condensed matter physics~\cite{amico2008,Schollwock2005}.    %It contains, more than many other measurable quantities,  detailed information on the wave function from which it is generated, without being overly complicated or hard to analyze.  Its importance in theoretical physics is multi-faceted with one of the most important aspects of its usage is that it directly determines the efficiency of representing a wave function as a matrix product state~\cite{Schollwock2005}.  
 An experimental measurement of quantum entanglement can provide information about the details of the wave functions for physical systems that we have yet to understand and would be of interest in many different systems, including Fermi liquids~\cite{McMinis2013,Swingle2013,Chandran2016,Junping2015,Gioev2006,Wolf2006}, topological phases~\cite{Levin2006,Zhuang2015,Kitaev2006,Zhang2011,Jiang2012,Zoller2016-1,Freeman2014} and even molecular systems~\cite{Tubman2012}.  %While there is no clear path to measuring entanglement in real materials, there are many interesting cold atom systems~\cite{McMinis2013,Chandran2016,Junping2015,Gioev2006,Wolf2006}.  
%For strongly correlated systems, such as the cuprates, it might provide a breakthrough for understanding the correlations of the many body wave function. 
% Unfortunately for realistic materials such measurements are not yet feasible.  However, in systems of cold atom experiments we describe how entanglement can be measured  by the latest advances in gas microscopes~\cite{Bloch2008,Ray2013,Mckay2015} with single site single particle resolution.  With such experimental setups it is feasible to measure quantum entanglement for a variety of Hamiltonians~\cite{Bakr2009}. 

 \begin{figure}[htpb]
\centering
\includegraphics[width=0.49\linewidth]{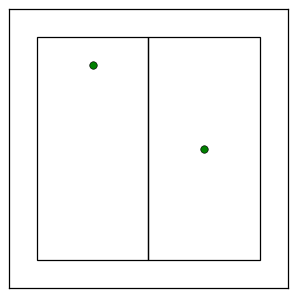}
\caption{Canonical example of where the fluctuation method fails to measure entanglement.  Two interacting atoms are separated by a divider that prevents the particles from fluctuating between both sides of the systems.  However, the particles are interacting with each other, and thus they have non-zero entanglement.  The fluctuation method fails to account for any of the quantum entanglement, however the information about their entanglement is encoded in the snapshots of a quantum gas microscope.  
}
\label{fig:entex} 
\end{figure}

The entanglement entropy, the Renyi entropy~\cite{amico2008}, and the entanglement spectrum~\cite{Li2008,Tubman2014}, are three of the most widely used quantities calculated from a spatial reduced density matrix.  %These quantities have been used to describe topological phase transitions~\cite{Zhuang2015}, many body localized states~\cite{Pal2010}, and establish a correspondence between condensed matter and string theory within the ADS/CFT formalism~\cite{Ryu2006}. 
 The spatial reduced density matrix is determined by splitting the system into two parts, regions A and B, where a density matrix in region A is given by integrating out all the degrees of freedom in region B.  This is written as follows,
\begin{equation}
\rho_{A}  = \textrm{Tr}_{B}(\rho_{AB}).
\end{equation}
The entanglement spectrum is defined as the eigenvalues of $\rho_{A}$, while the nth order Renyi entropy is given as 
\begin{equation}
S_{n}(A) = \frac{1}{1-n}\ln[\textrm{Tr}((\rho_{A})^{n})].
\end{equation}
The entanglement entropy is defined as 
\begin{equation}
S_{1}(A) = -\textrm{Tr}[(\rho_{A})ln(\rho_{A})].
\end{equation}

There are several techniques that have been proposed to measure quantum entanglement in cold atom systems~\cite{Klich2006,Abanin2012,Daley2012,Zoller2016}.  
Recent experiments have been able to realize one of these techniques~\cite{Islam2015}, however they were limited to a small number of particles and have not yet been able to address many predictions of interest~\cite{Gioev2006,Calabrese2004,Mahajan2016,Herdman2016,Tubman2012,Tubman2014,Tubman2015-1,Lundgren2016,Lundgren2016-1,Song2012} and disagreements such as the nature of the Widom conjecture in interacting systems~\cite{McMinis2013,Mishmash2016,Junping2015,Chandran2016,Chang2014}.
%, and are not in the realm of being able measure an area law or Widom conjecture violations~\cite{McMinis2013}.  %Generally we how to measure quantum entanglement in larger systems.
One of the techniques to measure quantum entanglement, which has been referred to as the \textit{the fluctuation technique}~\cite{Klich2006, Song2012} is derived from the properties of quantum entanglement of non-interacting systems.  Non-interacting systems have the special property that the entanglement entropy as well as the full entanglement spectrum can be determined by how particles fluctuate between two regions.  

\begin{figure}[htpb]
\centering
\includegraphics[width=0.49\linewidth]{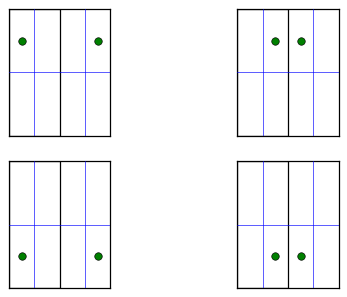}
\caption{Example of a highly entangled system of which there are no fluctuations between the half spaces.  There are two particles, each of which is constrained to be either in the left or right half of the system.  There are four lattice sites in each half.  Pictured are the only non-zero configurations of the wave function; each configuration has a 25\% probability.  For the two particles and four lattice sites, this systems has a large entanglement due to the mirror correlations.  The S$_{1}$ entropy is $-ln(0.25)$, which is equal to the largest entanglement two non-interacting particles can observe.  For lattices with more sites, the entanglement due to mirror correlations can increase past the limit for non-interacting particles.  A wave function with mirror correlations for 16 lattice sites in each half has an entanglement equal to $-ln(1/16)$.   The fluctuation method estimates a zero entanglement, where as the techniques discussed here have the potential to measure the exact entanglement. 
}
\label{fig:entex1} 
\end{figure}

A quantum gas microscope can be used to experimentally measure the fluctuations of particles in a cold atoms systems over any spatial partitions.  However, this technique can only serve as a lower bound of quantum entanglement when there are many body interactions.  The fluctuations make up only part of the entanglement, as both fluctuations and correlations are needed to determine the quantum entanglement~\cite{Klich2006, Song2012}. Examples of systems in which the entanglement is due entirely to many body correlations are illustrated in Fig. \ref{fig:entex} and Fig. \ref{fig:entex1}.  The system in Fig. \ref{fig:entex1}, which represents a wave function with mirror correlations, is one of the most strongly entangled wave functions that can be created. 
% In the previous section we introduced the no-nodes theorem for bosons, and demonstrated that a quantum gas microscope contains all the information needed to recreate the full wave function, which includes the quantum entanglement for any partitions of interest.
 It is important to emphasize that although the fluctuation method does not capture all the quantum entanglement for an interacting many body system
 %~\cite{Tubman2012,Song2012}, 
  the data from the quantum gas microscope  contains all the information needed to determine the  entanglement properties.
   
\prlsec{Swap Operator}
To understand how to measure entanglement of an interacting system we rely on the use of the swap operator which is an estimator used to calculate Renyi Entropies with quantum Monte Carlo~\cite{Hastings2010,Tubman2012}.  The operation of the swap operator is defined on a wave function that is written in terms of complete basis sets of regions A and B, $|\alpha \rangle$ and $|\beta \rangle$,  as follows $|\Psi\rangle = \sum C_{\alpha,\beta}|\alpha\rangle |\beta\rangle$.
The swap operator is defined in a space of the wave function that is a tensor product with itself as follows
\begin{eqnarray}
swap_{A}(\sum_{\alpha_{1},\beta_{1}}C_{\alpha_{1},\beta_{1}}|\alpha_{1}\rangle|\beta_{1}\rangle) \otimes
(\sum_{\alpha_{2},\beta_{2}}D_{\alpha_{2},\beta_{2}}|\alpha_{2}\rangle|\beta_{2}\rangle) \nonumber \\
= \sum_{\alpha_{1},\beta_{1}}C_{\alpha_{1},\beta_{1}}
\sum_{\alpha_{2},\beta_{2}}D_{\alpha_{2},\beta_{12}}(|\alpha_{2}\rangle|\beta_{1}\rangle
\otimes |\alpha_{1}\rangle|\beta_{2}\rangle ).\nonumber \\
\end{eqnarray}

The expectation value of the swap operator is related to $\rho_{A}$ by the following,
\begin{equation}
 Tr((\rho_{A})^{2}) = \frac{\langle \Psi_{T} \otimes \Psi_{T}| swap_{A} |\Psi_{T} \otimes \Psi_{T} \rangle}{\langle \Psi_{T} \otimes \Psi_{T}|\Psi_{T} \otimes \Psi_{T} \rangle}.
\end{equation}
  This operator can be sampled with independent configurations generated from $|\Psi_{T}|^{2}$ and the results can be used to calculate the Renyi entropy, S$_{2}$ = $-ln(Tr(\rho_{a})^{2})$.
In terms of the real space coordinates the expectation value of the swap operator is  given by Eq. \ref{eqn:estimator}.
\begin{figure*}
\begin{eqnarray}
\label{eqn:estimator}
 \langle \Psi_{T} \otimes \Psi_{T}| swap_{A} |\Psi_{T} \otimes \Psi_{T} \rangle
= \int d\mathbf{x}_{1}d\mathbf{x}_{2} \cdots d\mathbf{x}_{2N} \Psi^{*}_{T}(R(\alpha_{1},\beta_{1}))\Psi^{*}_{T}(R(\alpha_{2},\beta_{2}))\Psi_{T}(R(\alpha_{2},\beta_{1}))\Psi_{T}(R(\alpha_{1},\beta_{2})) \nonumber \\
= \int d\mathbf{x}_{1}d\mathbf{x}_{2} \cdots d\mathbf{x}_{2N} |\Psi_{T}(R(\alpha_{1},\beta_{1}))|^{2}|\Psi_{T}(R(\alpha_{2},\beta_{2}))|^{2}\frac{\Psi_{T}(R(\alpha_{2},\beta_{1}))\Psi_{T}(R(\alpha_{1},\beta_{2}))}
{\Psi_{T}(R(\alpha_{1},\beta_{1}))\Psi_{T}(R(\alpha_{2},\beta_{2}))}
\end{eqnarray}
\end{figure*}
From these definitions the operator can be identified as
\begin{equation}
O(\alpha_{1},\alpha_{2},\beta_{1},\beta_{2}) = \frac{\Psi_{T}(R(\alpha_{2},\beta_{1}))\Psi_{T}(R(\alpha_{1},\beta_{2}))}
{\Psi_{T}(R(\alpha_{1},\beta_{1}))\Psi_{T}(R(\alpha_{2},\beta_{2}))}
\label{vmcop}
\end{equation}
In these equations we have used the notation $R(\alpha,\beta)$, which are the real space coordinates.  The swap operator is performed over two sets of  coordinates, and thus the subscripts are used to identify which set of coordinates to use.  All the coordinates in region A are associated with $\alpha$ and all the coordinates in region B are associated with $\beta$. 

Once $\psi_{exp}$ has been generated, either through full wave function tomography, or correlation function matching, it is then straightforward to generate new configurations using $\psi_{exp}$, and evaluating the estimator in Eq.~\ref{vmcop}, by replacing $\psi_{T}$ with $\psi_{exp}$.  We have tested this process in previous works which includes Fermi liquids and molecular systems~\cite{Tubman2012,McMinis2013,Swingle2013}, and other authors have done related studies on systems of bosons~\cite{Herdman2014,Herdman2015}.   Further possibilities of evaluating the full entanglement spectrum are possible as described in other QMC work~\cite{Tubman2014,Chung2013}.

\prlsec{From Theory to Experiment: Phase identification with Neural Networks and Density functional theory}
In this section we consider going in the other direction, in which theoretical tools are first developed and then applied to quatum gas microscope data.  %, and %a tool is generated that can be used to identify the phase of a state given a series of snapshots.   %A precise form of this has been given previous under certain circumstances, and here we figure out how to generalize this further.
Recent developments in artificial neural networks have shown that phases can be learned from quantum Monte Carlo snapshots of a system~\cite{Melko2016-3}. %which is a different idea than using a neural network with symmetry functions to generate boson wave functions as described in the previous section. 
 The idea is to generate quantum Monte Carlo configurations from different phases that are known theoretically and train a neural network to recognize the different phases.  Once trained the neural network serves as a tool to identify quantum phases, which can be applied to the data from a quantum gas microscope.   %  The next step is to take such a neural network and apply it to the data of a quantum gas microscope to get the best guess of what phase a system is observing.  
  In recent works~\cite{Melko2016-1,Melko2016-2,Melko2016-3} the details of how to do some phase discrimination has been discussed in detail.

Advances for applying these machine learning techniques to fermions has produced promising results.  The original technique~\cite{Melko2016-1} and has since been adapted to Green's functions for Fermion systems~\cite{Melko2016-2}, and recently to more general data~\cite{Melko2016-3}.  There are tradeoffs of such an approach compared to what would be a more straightforward calculation of an order parameter, or in comparison to the tomographic methods described in this work.  Advantages include that one does not have to know the order parameter to begin with, it works inherently at zero temperature and finite temperature, and one does not even have to know what the Hamiltonian for the system in which data is being generated.  The negatives are that one does not know how transferable learning between different Hamiltonians or even between data at different temperatures.  Neural networks should be considered good at interpolation and bad at extrapolation.  If a phase of matter has not been included in a training set, then it is unlikely a neural network will produce meaningful results when asked to identify such a phase.  
In comparison to the wave function tomography presented in this work, some of the features are similar.  A finite temperature tomography requires density matrix optimization~\cite{Blunt2014,Baumgratz2013,Ceperley1995}, which is possible but requires one to extend the techniques presented in this work.    
For the wave function tomography, one must include enough variational degrees of freedom to learn a wave function properly, where as with the neural networks, one has to train on all possible phases that a new system might observe.  In both cases some care is required before applying the techniques.

%In a similar manner as to the wave function tomography and phase learning through neural networks,
We also note that  density functional theory for cold atom systems has recently been developed for the purposes of theoretical modeling~\cite{ma2012}.  This approach can also be applied to gas microscope data of cold atom systems.  The one particle correlation function is the density of a system which can be directly estimated from a quantum gas microscope.  In the same manner as described for the neural networks, one can use the best functional developed for a system of interest, and map out the phase diagram with energies generated from density functional theory.  Thus one can test different approaches ranging from wave function tomography, DFT analysis, and phase learning from neural networks to get a very detailed picture of a system of interest.

%\prlsec{Fitting Correlation functions}
%\prlsec{Data/Discussion}
\prlsec{Conclusions}
We presented several ideas for studying quantum gas microscope data with the main goal of demonstrating that such data can be used for full wave function tomography on a wide range of quantum systems.  %In this work we presented an extensive approach to analyzing the data of a quantum gas microscope  
%There are many connections between theory and experiment that be made through modern computational techniques and recent experimental advances with the quantum gas microscope.  %We show that the data from a quantum gas microscope can be used to get a very detailed picture of a quantum wave function through either wave function optimization or correlation function matching.   
The wave function tomography approaches facilitate experimental measures of what are generally hard quantities to measure, such as the quantum entanglement.  The two techniques presented here can be applied in different scenarios related to the amount of data that can be generated in an experiment.  Both techniques take advantage of the fact that many body correlations and fluctuations are encoded in the quantum gas microscope snapshots.  With this analysis future quantum gas microscope experiments should be able to elucidate many of the exciting features that are hidden within a quantum wave functions.   %and in the case of boson wave functions, the entire wave function is actually encoded.

\section{Acknowledgements}
We would like to thank Ushnish Ray, Daniel Freeman, Bill Huggins, Miles Stoudenmire, Birgitta Whaley and Brian Swingle for useful discussions.
This work was supported through the Scientific Discovery through
Advanced Computing (SciDAC) program funded by the U.S. Department of
Energy, Office of Science, Advanced Scientific Computing Research and
Basic Energy Sciences. We used the Extreme Science and Engineering Discovery
Environment (XSEDE), which is supported by the National Science Foundation Grant No. OCI-1053575 and
resources of the Oak Ridge Leadership Computing Facility (OLCF) at the Oak Ridge National Laboratory,
which is supported by the Office of Science of the U.S.
Department of Energy under Contract No.  DE-AC0500OR22725.

%\begin{figure*}[htpb]
%\centering
%\includegraphics[width=1\linewidth]{./nmer_V.pdf}
%\caption{(Color online): Central cut entanglement entropy for 
%$n$-mer models ($n$=1,2,3,4) for various fillings and disorder strengths. 
%}
%\label{fig:nmer_V} 
%\end{figure*}	

%\bibliography{refs}{}

%

\end{document}